\documentclass[runningheads]{llncs}
\usepackage{latexsym,amsmath}
\usepackage{amssymb}

\begin{document}


\title{Linear Non-Transitive Temporal Logic, Knowledge Operations, Algorithms for Admissibility }



\author{Vladimir Rybakov}
\authorrunning{V.Rybakov}
\titlerunning{Agent's Knowledge Operations in Non-Transitive Temporal Logic
 }

\institute{School of Computing, Mathematics and DT,
  Manchester Metropolitan University,
 John Dalton Building, Chester Street, Manchester, M1 5GD, U.K,
 \email{V.Rybakov@mmu.ac.uk}
 }

\toctitle{Lecture Notes in Computer Science}

\mainmatter

\maketitle

\newcommand{\ga}{\alpha}
\newcommand{\gb}{\beta}
\newcommand{\grg}{\gamma}
\newcommand{\gd}{\delta}
\newcommand{\gl}{\lambda}
\newcommand{\cff}{{\cal F}or}
\newcommand{\ca}{{\cal A}}
\newcommand{\cb}{{\cal B}}
\newcommand{\cc}{{\cal C}}
\newcommand{\cm}{{\cal M}}
\newcommand{\cmm}{{\cal M}}
\newcommand{\cbb}{{\cal B}}
\newcommand{\ccrr}{{\cal R}}
\newcommand{\cf}{{\cal F}}
\newcommand{\cy}{{\cal Y}}
\newcommand{\cxx}{{\cal X}}
\newcommand{\cdd}{{\cal D}}
\newcommand{\cww}{{\cal W}}
\newcommand{\czz}{{\cal Z}}
\newcommand{\cll}{{\cal L}}
\newcommand{\cw}{{\cal W}}
\newcommand{\ckk}{{\cal K}}
\newcommand{\ppp}{{\varphi}}

\newcommand{\ii}[0]
{\rightarrow}

\newcommand{\ri}[0]
{\mbox{$\Rightarrow$}}

\newcommand{\lri}[0]
{\mbox{$\Leftrightarrow$}}

\newcommand{\lr}[0]
{\mbox{$\Longleftrightarrow$}}

\newcommand{\ci}[1]{\cite{#1}}

\newcommand{\pr}{{\sl Proof}}

\newcommand{\vv}[0]{
\unitlength=1mm
\linethickness{0.5pt}
\protect{
\begin{picture}(4.40,4.00)
\put(1.2,-0.4){\line(0,1){3.1}}
\put(2.1,-0.4){\line(0,1){3.1}}
\put(2.1,1.1){\line(1,0){2.0}}
\end{picture}\hspace*{0.3mm}}}

\newcommand{\nv}[0]{
\unitlength=1mm
\linethickness{0.5pt}
\protect{
\begin{picture}(4.40,4.00)
\put(1.2,-0.4){\line(0,1){3.1}}
\put(2.1,-0.4){\line(0,1){3.1}}
\put(2.1,1.1){\line(1,0){2.0}}
\protect{
\put(0.3,-0.7){\line(1,1){3.6}}}
\end{picture}\hspace*{0.3mm}
}
}

\newcommand{\nvv}[0]{
\unitlength=1mm
\linethickness{0.5pt}
\protect{
\begin{picture}(4.40,4.00)
\put(2.1,-0.4){\line(0,1){3.1}}
\put(2.1,1.2){\line(1,0){2.0}}
\protect{
\put(0.6,-0.5){\line(1,1){3.6}}}
\end{picture}\hspace*{0.3mm}
}
}

\newcommand{\dd}[0]{
\rule{1.5mm}{1.5mm}}

\newcommand{\llll}{{\cal LT \hspace*{-0.05cm}L}}

\newcommand{\nn}{{\bf N}}

\newcommand{\pp}{{{\bf N^{-1}}}}

\newcommand{\uu}{{{\bf U}}}

\newcommand{\suu}{{{\bf S}}}

\newcommand{\bb}{{{\bf B}}}

\newcommand{\nnn}{{\cal N}}
\newcommand{\sss}{{{\bf S}}}
\newcommand{\zzz}{{\cal LT \hspace*{-0.05cm}L}_K(Z)}
\newcommand{\zz}{{{\cal Z}_C}}

\date{}

\bigskip

\begin{abstract} The paper studies problems of satisfiability, decidability and
admissibility of inference rules, conceptions of knowledge and agent's knowledge in
non-transitive temporal linear logic $LTL_{Past,m}$. We find algorithms solving mentioned
problems, justify our approach to consider linear non-transitive time with several examples.
 Main, most complicated, technical new result is {it \bf decidability of $LTL_{Past,m}$ 
 w.r.t. admissible rules}.
We discuss several ways to formalize conceptions of knowledge and agent's knowledge
within given approach in non-transitive linear logic with models directed to {\em past}.
     \end{abstract}



\  parameterized knowledge operations,
satisfiability, admissible
 rules, deciding

\ algorithms


\section{Introduction}

No question that there is no more interesting and mysterious
object as conception of time.
But nowadays it works fine in prose of the life -- for example, it has  many
interpretations in CS   (eg.
for interpretation of spread the set of check points in computational runs, etc.).
 Historically, investigations of temporal logic in mathematical/philosophical logic
 based at modal systems was originated by Arthur Prior in the late 1950s.
 Since then temporal logic has been (and is) very active area in mathematical logic and information sciences, AI
 and CS (cf.  eg. --   Gabbay and Hodkinson\cite{ghr,gbho,gbho2}).
It was observed that temporal logic has
important applications in formal verification, where it is used to state requirements of hardware or software systems.

In particular, linear temporal logic $ \llll $ (with Until and Next)
is very useful instrument(cf.
Manna, Pnueli \ci{ma1,ma2}, Vardi \ci{va1,va2}) ($\llll$ was used for analyzing protocols of computations, check of consistency, etc.). The decidability and satisfiability problems for $\llll$, so to say main problems,
were in focus of investigations  and were successfully  resolved (cf. references above).

The conception of knowledge, and especially the one implemented via multi-agent approach is a popular area in Logic in Computer Science. Various aspects including interaction and autonomy, effects of cooperation etc
were investigated (cf. eg. Woldridge et al  \cite{wl1,wl2,wl3}, Lomuscio et al \cite{lo1,be11}). In particular, a multi-agent logic with distances were studied and satisfiability problem for it was solved (Rybakov et al \cite{ry10t});
conception of Chance Discovery in multi-agent's environment was considered (Rybakov \cite{ry11a,r12t});
 a logic modeling uncertainty via agent's views was investigated (cf. McLean et al \cite{mcln1});
representation of agents interaction (as a dual of common knowledge) was suggested in Rybakov \cite{ry09t,vr179}.

Conception of refined common knowledge was suggested in Rybakov \cite{ry2003}. Historically the conception of
common knowledge was formalized and profoundly analyzed in 1990x, cf. eg. Fagin et al \cite{fag1}, using as a base
agent's knowledge (S5-like) modalities. The approach to model knowledge in  terms of symbolic logic, probably, may be dated to the end of 1950.
At 1962 Hintikka \cite{jh4} wrote the book: {\em Knowledge and Belief}, the first book-length work to suggest using modalities to capture the semantics of knowledge.

In contemporary study, the field of knowledge representation and reasoning in logical terms is very wide and active area.
Frequently modal and multi-modal logics were used for formalizing agent's reasoning.
Such logics
were, in particular, suggested in Balbiani et al \cite{pb1},
Vakarelov \cite{dv},
Fagin et al
\cite{fag1}, Rybakov \cite{ry2003,vr179}. The book Fagin et al
\cite{fag1} contains summarized to that time systematic approach to study
the notion of common knowledge.
Some contemporary study of knowledge and believes in terms of single-modal logic may be found in
  Halpern et al \cite{hal1}.
Modern approach to knowledge frequently uses
conception of justification in terms of epistemic logic
(cf. eg. Atremov et al \cite{art1,art2}, Halpern \cite{hal1}). We will suggest some views on knowledge and knowledge in terms of multi-agent logic being based at refined version of linear temporal logic $LTL$. We will need some technique borrowed from tools for verification of admissibility for inference rules.

The problem of admissibility for inference rules and unifiability problem were already addressed to
linear temporal logic.
The admissibility problem (to determine for any given rule if this
rule is admissible for a given logic) was in focus of interest for
many logicians. Active research in the area may be dated to Harvey
Friedman problem \cite{fr}: if there is an algorithm of verification
for admissibility in the intuitionistic propositional logic
$\mathbf{IPC}$ (this problem was first solved by Rybakov in 1984,
\cite{ry84}). Since then many logicians were interested to study
admissibility from various viewpoints and in many logical systems
related to non-classical propositional logics
 (cf. V.~Rybakov
\cite{r4,Rybakov92,Rybakov2001,r6,r7,ry08a}, Rybakov, et al \cite{babry,babry1},
R.~Iemhoff \cite{r1,r3},
 R.~Iemhoff and G.~Metcalfe \cite{r322},
 E.~Jerabek \cite{r61,r62,r63,r61a}); prime questions were recognizing admissibility, study of bases for inference rules.
 Only a necessary condition for admissibility of inference rules in the branching-time temporal logic $T_{S4}$ was found in \cite{ry08aa}, though for linear temporal logic LTL the problem was solved in  full \cite{ry08a}.
  Complexity problem for admissibility in intuitionistic logic and some modal logics was first studied and
  re-solved in Jerabek \cite{r61bb}.

Effective  approach to study admissible rules was offered by S.~Ghilardi
via unification technique; at (\cite{ghi1}, 1999) it was first found and algorithm
writing out a complete set of unifiers for any unifiable in $\mathbf{IPC}$ formula, and this
gives another solution for admissibility problem.
Since then, unification in propositional modal logics
over K4
was extensively studied by S.~Ghilardi~\cite{ghi1,ghi2,ghi10}. He
developed a novel method, based on L\"owenheim approach, which has
proved to be also useful in dealing with admissibility and bases of
admissible rules.
  In (Babenyshev and Rybakov \cite{babry}) we found solution of the unification problem in the linear temporal logic LTL.
(the case of $\llll$ with no Until was solved; the case of linear temporal logic with future and past easy follows because we may model in this logic the universal modality (cf. Rybakov \cite{r7})).

Our paper studies a non-transitive version of the linear temporal logic $\llll$ -
the linear temporal logic $LTL_{Past,m}$ based at non-transitive linear frames.
We consider the frames diverted to {\em past} and since that we  as the basic operation
{\em Since} instead of {\em Until}.
Based at the property of operation {\em Since} (to keep
truth values of a statements since a specified  property happened to be true)
we define a conception to be {\em knowledge} in several possible ways including {\em parameterized knowledge}
and {\em knowledge via agent's viewpoint, including voting, preceding events etc.}.
We find this to be very plausible and useful interpretation.

We begin form construction the logic $LTL_{Past,m}$: we define its language, syntax, semantic models.
Then we shortly comment why these models look plausible and give several examples.
Next, we develop all necessary technical instruments, including notion of valid inference rules, and reduced normal forms for inference rues. Based at this, we solve satisfiability and decidability problems for
$LTL_{Past,m}$ (these results 
 similar to the ones  submitted to a conference  \cite{vr14g}, for the case non-uniform bounds of intransitivity). 
 Then we approach to the main problem solved in this paper: we show that $LTL_{Past,m}$ is decidable w.r.t.
admissible rules and we find an algorithm solving admissibility problem.
This is the main (and most complicated) new technical result of this paper.
We, in the concluding  part,  comment how this approach might be used in interpretation of logical knowledge and agent's knowledge operations.
The paper is about self contained and  use only technique which is  explained and defined here
(proofs however due to length are usually omitted).



\section{Necessary preliminary information}

We will base our approach at a new non-transitive version of the linear temporal logic $\llll$
(motivation for non-transitivity will be given at short separate section below).
Therefore we start from a sort recall of notation and definitions concerning $\llll$ .
The language of the Linear Temporal Logic ($ \llll $ in the sequel)
extends the language of Boolean logic by operations $\nn$ (next)
and $\uu$ (until). The formulas of $\llll$ are built up from a set $Prop$ of atomic
 propositions (synonymously - propositional letters)
  and are closed under applications of Boolean
 operations, the unary operation $\nn$ (next) and the binary
 operation $\uu$ (until). The formula $\nn\ppp$ has meaning: the statement $\ppp$ holds in the next time
point (state); the formula $\ppp \uu \psi$ means: $\ppp$ holds until
$\psi$ will be true. Semantics for $\llll$ consists of {\em infinite
  transition systems (runs, computations)}; formally  they are
  represented as
  linear Kripke structures based on natural numbers.
The infinite linear Kripke structure is a
 quadruple \(
\cm:=\langle \nnn, \leq, \mathrm{Next}, V
 \rangle,\)
 where $\nnn$ is the set of all natural
 numbers;   $\leq$ is the
  standard order on $\nnn$, $\mathrm{Next}$ is the binary relation,
  where $a \ \mathrm{Next} \ b$ means $b$ is the number
 next to $a$. $V$ is a valuation of a subset $S$ of
 $Prop$.  Hence the valuation  $V$
 assigns truth values to elements of $S$.

 So, for any
 $p\in S$, $V(p)\subseteq \nnn$, $V(p)$ is the
 set of all $n$ from $\nnn$
 where $p$ is true (w.r.t. $V$).
All elements of $\nnn$ are called to be {\em states} (worlds), $\leq$ is the {\em
transition} relation (which is linear in our case), and $V$ can be
interpreted as {\em labeling} of the states with atomic
propositions.
The triple $\langle \nnn, \leq, \mathrm{Next}
 \rangle$ is a Kripke frame which we will denote for short by
 $\nnn$.

The truth values in any Kripke structure
 $\cm$,
can
be extended from propositions of $S$ to arbitrary formulas
constructed from these propositions as follows:

\smallskip

\begin{definition} Computational rules for logical operations:

\begin{itemize}

\item $ \forall p\in Prop \ (\cm,a)\vv_V p \   \lri   \ a\in \nnn \wedge \ a\in V(p);$

\medskip

\item $ (\cm,a)\vv_V (\ppp\wedge \psi)  \ \lri \ $
$(\cm,a)\vv_V \ppp \wedge
(\cm,a)\vv_V \psi;$

\medskip

\item $ (\cm,a)\vv_V \neg \ppp \   \lri  \ not [(\cm,a)\vv_V \ppp]  ;$

\medskip

\item  $ (\cm,a)\vv_V \nn  \ppp \  \lri  [\ \forall b
[(a \ \mathrm{Next} \ b) \ri (\cm,b)\vv_V \ppp]] ;$

\medskip

\item $ (\cm,a) \vv_V (\ppp \uu \psi)  \ \lri  \ \exists b [
(a\leq b)\wedge ((\cm,b)\vv_V \psi) \wedge $
\item $
 \ \ \ \ \ \ \ \ \ \ \forall c [(a\leq c < b) \ \  \ri \ \  (\cm,c)\vv_V \ppp ]].
$

\end{itemize}

\end{definition}

\smallskip



 { For a Kripke structure $\cm:=\langle \mathcal{N}, \leq,
\mathrm{Next},V
 \rangle$ and a formula $\ppp$ with letters from the domain of $V$,
  we say
$\ppp$ is valid in $\cm$ (denotation -- $\cm\vv \ppp$) if, \ for any
$b$ of $\cm$ ($b\in \mathcal{N}$),
 the formula $\ppp$ is true at $b$ (denotation: $(\cm,b)\vv_V
\ppp)$.}

The linear temporal logic $\llll$ is the set of all formulas
  which are valid in all infinite temporal linear Kripke structures
  $\cm$ based on $\nnn$ with standard $\leq$ and $\mathrm{Next}$.
  Now we will modify the models for the ones with non-transitive  time.

 \section{Possible words models with non-transitive time}

 Our approach  will  need a dual of $\llll$  -- the logic with {\em since} operation.
 Actually we may interpret this dual  is as a standard LTL, but with time diverted to past,
 and yet based at non-transitive models.
We may introduce this dual as follows.
The formulas are constructed as earlier,
but with the binary logical operation $\suu$ (since) instead of $\uu$ (until).


The frame $\nnn^{-}$ is $\langle  N,  \geq, \mathrm{Next}  \rangle$,
 and $V$ as before is a valuation of a subset $S$ of
 $Prop$ on the set $N$. So, we  take the language of $\llll$, delete $\uu$ and replace it with
   the binary operation $\suu$. The definition of the truth relation for
 $\suu$ is as follows:

\[ (\nnn^{-},a) \vv_V (\ppp \suu \psi) \ \lri \exists b [
(b\geq a)\wedge ((\nnn^{-},b)\vv_V \psi) \wedge \]

\[
 \hspace*{0.1cm}\forall c [(a \leq c < b) \ri (\nnn^{-},c)\vv_V \ppp ]].
 \]
So, $\suu$ is just the dual of $\uu$ (and note that it acts exactly as $\uu$, we simply interpret it to {\em past}).
Operation {\em Next} (notation $\nn$) will act as earlier, but again directed to the past, {\em next one means next in past}. That is

\[ (\nnn^{-},a) \vv_V  \nn \ppp \ \lri \  (\nnn^{-}, a+1)\vv_V \ppp. \]
(We will not use standard notation for LTL with {\em until} because it will break our approach
to model knowledge in later part in this paper.)

\begin{definition}
{\it \bf  A non-transitive possible-worlds linear frames with uniform non-transitivity} (which upon we will base our approach)
is a freame:

\[\cf := \langle N, \geq, \mathrm{Next}, \bigcup_{ i \in N} R_i  \rangle,\]
where each $R_i$ is the standard linear order ($\geq$) on the interval $[i,i+m]$, where
$m$ is a fixed natural number (measure of intransitivity).
\end{definition}

For any set of letters $P$ we may define an arbitrary valuation $V$ on $\cf$ in standard
way, and $\cf$ with a valuation is called a model $\cm$. Thus, for any $p \in P$, 
we have $V(p) \subseteq N$ and 
we may extend $V$
 to all boolean formulas built up from $P$ as usual. The same way as earlier we define truth
values for formulas of kind $ \nn \ppp$.
But for formulas  $ \ppp \suu \psi$ the definition is new one:

\begin{definition} Computation rule for {\bf weak bounded since}:

\smallskip

$$(\cm,a) \vv_V (\ppp \ \suu \ \psi) \  \ \ \lri  \ \ \  $$

$$ \exists b [
(b R_{a}   a)\wedge ((\cm,b)\vv_V \psi) \wedge
\forall c [(a \leq c < b) \ri
(\cm,c)\vv_V \ppp ]].
 $$
 \smallskip
\end{definition}

\begin{definition}
The logic $LTL_{Past,m}$ is the set of all formulas which are valid  at any model $\cm$ with the measre of intransitivity
$m$.
\end{definition}

The relation
$\bigcup_{ i \in N} R_i $ is evidently non-transitive and right now we
will explain why this approach to time is accepted. Just to immediately comment,  briefly note  that
$\suu$ works similar to usual
$\uu$ (but only within the stet of states accessible from the current one,
-- since our accessibility relations are non-transitive); and yet it
acts as {\em Since} regarding our definition of the time accessibility relation directed to past.

\section{Why we consider that time might be non-transitive}

The approach described above considers the time directed to past (which, so to say, relays
to our memory about past). Computationally (form CS viewpoint), we analyze the behavior of
a computation, computational runs, being given by protocols of events which happened
(already happened) while computation.

\smallskip

{\bf View (i)}. { \em Computations view}. Inspections of protocols for computations are limited by time resources and
have non-uniform length. Therefore, if we interpret our models as the ones reflecting inspection of
protocols for computation, the amount of check points is finite. In any point of inspection we may
refer to stored protocols, and any one has limited length. Thus the inspections look as non-transitive accessibility relations.

\smallskip

{\bf View (ii)}. {\em Agent's-admin's view.}
We may consider states (worlds of our model) as checkpoints of admins (agents) for any
inspection of state of network in past.
Any admin has allowed amount of inspections for previous  states,
but only within the areas of its(his/her) responsibility  (by security or another reasons).
So, the accessibility is not transitive again, the admin (a1) can reach a state, and there in, the admin (a2) responsible for this state (it may be a new one or the same yet),
has again some allowed amount of inspections to past.
But, in total, (a1) cannot inspect all states accessible for (a2).

\smallskip

{\bf View (iii)}. {\em Agent's-users's view.} If we consider the sates of the models as the content of web pages admissible for users, any surf step is accessibility relation, and starting from any web page user may achieve, using links in hypertext(s)  some foremost available web site. The latter one may have web links which are available only for individuals possessing passwords for accessibility. And these one (having password) may continue web surf, etc. Clearly that in this approach,  web surfing looks as non-transitive relation. Here we interpret web surf as time, but diverted to past, instead
of opposite. The models suggested above again serve well these approach.

\smallskip

{\bf View (iv)}. {\em View on time in past for collecting knowledge.}
In human perception, the some finite time in past (not in future)
is only available to individuals to inspect evens and knowledge collected to current time state.
The time is past in our feelings looks as linear, and, as an individual,
the same as mankind in total, have finite amount
of memory to remember information and events.
There, in past, at foremost available (memorable) time point, individuals again had a memorable interval of time with collected information. And so forth...
So, the time in past for mankind overall is not transitive.

\smallskip

{\bf View (v)}. {\em View in past for individuals as agents with opposition.}
Here the picture is similar  to the case (iv) above, but we may consider the knowledge as
the collection of facts which about majority experts (agents) have affirmative positive opinion.
And, in past time, the voted opinion of experts about facts could be different at
distinct time points, Besides the time intervals memorable by experts might be very variative in past.
Therefore in this approach time nohow  may look transitive.

\section{Satisfiability,  decidability, admissible rules, computation algorithms, }

Since we introduced a sensibly innovated linear temporal logic (modification of LTL, a very popular in CS logic), which is non-transitive, we would like
to address first to this logic basic computational problems for any logic: satisfiability and decidability problems.
It is immediately seen that the old standard techniques to solve these problems do not
work because the accessibility relation is not transitive.
Therefore the standard technique of rarefication as eg. at \cite{ry09t,vr179} or direct usage of automatons technique do not work here.
If we would try to use some variants of filtration technique
than a technical obstacle is possible nested operations $\suu$ and non-transitivity again: how we would  define
accessibility relations $R_i$ then.

Therefore we will need some preliminary work to avoid nested operations.
In fact we will be based on a modernization of the technique
used already, eg. in
\cite{ry09t},
 Most gain from this approach is that we will
then consider only very simple and uniform formulas which are formulas
without nested temporal operations. For this, we will need  a transformation of formulas into rules
in  reduced form. Recall that
a (sequential) (inference) rule is
 an expression (statement)

 \[ {\bf r}:= \frac{\varphi_1( x_1, \dots ,
x_n), \dots , \varphi_l( x_1, \dots , x_n)}{\psi(x_1, \dots , x_n)},
\]
where $\varphi_1( x_1, \dots , x_n), \dots , \varphi_l( x_1,
\dots , x_n)$ and $\psi(x_1, \dots , x_n)$ are
 formulas constructed out of
letters $x_1, \dots , x_n$. The letters $x_1, \dots , x_n$ are the
variables of ${\bf r}$, we use the notation $x_i \in Var(\bf r)$. A
meaning of a rule {\bf r} is that the statement (formula) $
\psi(x_1, \dots , x_n)$ (which is called conclusion) follows
(logically follows) from statements (formulas) $\varphi_1( x_1,
\dots , x_n),$  $\dots ,$  $\varphi_l( x_1, \dots , x_n)$ which are
called premisses.

\begin{definition} {\it
A rule
  ${\bf r}$ is said to be {\em valid} in
a  model $\langle { \cm }, V \rangle$
(we will use the notation ${\cm}\vv_V \ {\sl
r}$) if

\[ [\forall a \ (({\cm,a}) \vv_V \bigwedge_{1\leq i \leq
l}\varphi_i)] \Rightarrow [ \forall a \ (({\cm},a) \vv_V \psi)].\]
Otherwise we say ${\bf r}$ is {\sl refuted} in $\cm$, or {\sl
refuted in $\cm$ by $V$}, and write ${\cm }\nv_V {\bf r}$. A rule
${\bf r}$ is {\sl\ valid } in a frame ${\cm}$ (notation ${\cm}\vv \
{\bf r}$) if, for any valuation $V$, the following holds ${\cm}\vv_V {\bf r}$}.
\end{definition}

For any formula $\ppp$,  we can transform $\ppp$ into the
  rule $x \ii x /
\ppp$ and employ a technique of reduced normal forms for inference
rules as follows.

\begin{lemma} \label{p1} {For any formula $\ppp$, $\ppp$  is a theorem of $LTL_{Past,m}$
iff the rule $({x \ii x / \ppp})$ is valid in any frame  ${\cm}$ .}
\end{lemma}

\begin{definition}
 A rule ${\bf r}$  is said to be in
{\em reduced normal form} if \( {\bf r}= \varepsilon / x_1\) where

\[\varepsilon :=
 \bigvee_{1\leq j \leq l}
 [ \bigwedge_{1\leq i \leq n}
 x_i^{t(j,i,0)} \wedge
\bigwedge_{1\leq i \leq n}
 (\nn x_i)^{t(j,i,1)} \wedge
\bigwedge_{1\leq i,k \leq n, i \neq k}
(x_i \suu x_k)^{t(j,i,k,1)}] \]
   and,
 for any formula $\ga$ above,
 $\ga^0 := \ga$, $\ga^1:= \neg \ga$.
\end{definition}

\begin{definition} {\it
Given a rule ${\bf r_{nf}}$ in  reduced normal form, ${\bf r_{nf}}$
is said to be a {\it normal reduced form for a rule ${\bf r }$} iff,
for any frame ${\cal F}$ for $LTL_{Past,m}$,

\[ {\cal M} \vv {\bf r} \ \lri {\cal M}  \vv {\bf
r_{nf}}  .\]}
\end{definition}

\begin{theorem} \label{mt3} {\it There exists
an algorithm  running in (single) exponential time, which, for any
given rule ${\bf r}$, constructs its normal reduced form ${\bf
r_{nf}}$}.
\end{theorem}

Now we need we need special
finite frames $\cf(N^{-,m})$ having the structure resembling
the structure of frames $\cf$ but not linear discrete ones (which are not the frames which
we used to define
 $LTL_{Past,m}$). We have no space to define their structure explicitly due to paper space limitation.
Note only that the accessibility relations at
frames $\cf(N^{-,m})$ are non-transitive and with measure of transitivity $m$ also.

\begin{lemma} \label{oo1} {\sl For any given  rule $\bf r_{nf}$ in  reduced
normal form, $\bf r_{nf}$  is refuted  in a frame of $\cf$ iff $\bf r_{nf}$ can be
refuted in some finite frame $\cf(N^{-,m})$ by a valuation $V$, where the size
of the frame $\cf(N^{-,m})$ has size effectively computable from the
size of $\bf r_{nf}$. }
\end{lemma}

It is clear that a formula $\ppp$ is satisfiable ($\ppp$ is true at a state of a model for $LTL_{Past}$)
if and only if $\neg \ppp \not\in LTL_{Past,m}$.
Therefore based at  Theorem \ref{mt3}, Lemma \ref{p1} and Lemma \ref{oo1} we
may prove:

\begin{theorem} \label{bn1} {\it The logic
 $LTL_{Past,m}$ is
decidable; the satisfiability problem for the non-transitive linear temporal logic
 $LTL_{Past,m}$ is
decidable: for any formula we can compute if it is satisfiable and
to compute the valuation of the model satisfying this formula.}
\end{theorem}

The computational algorithm, in its final stage, consists of definition a computable valuation in some initial part of $N$
which finally is resulted in a total valuation satisfying the formula.
Now on we have enough technique to proceed to admissibility of inference rules.

\begin{definition}{\em
An inference rule
 
 \( {\bf r}:= {\varphi_1( x_1, \dots ,
x_n), \dots , \varphi_l( x_1, \dots , x_n)} / {\psi(x_1, \dots , x_n)},
\)
 is said to be 
 
 {\em admissible }  
 in a logic $L$ if, for any tuple of formulas
 $\ga_1 , \dots , \ga_n $, the following 
 
 holds
 \(  [ \bigwedge_{1\leq
i \leq l} \varphi_i(\ga_1, \dots , \ga_n)\in L]
 \ \ \Rightarrow \ \ 
[ \psi(\ga_1, \dots , \ga_n) \in L ]. \)
}
\end{definition}

 Thus, for any admissible rule,
 any instance into the premises
making all of them theorems of a logic $L$  makes also the conclusion to
be a theorem. Using the same algorithm of construction reduced normal form
$\bf r_{nf}$ for any given rule $\bf r$ as at Theorem $\ref{mt3}$ we may
obtain

\begin{lemma} \label{ooxx} {\sl
For any given  rule $\bf r$,
$\bf r$ is admissible in $LTL_{Past,m}$ iff
$\bf r_{nf}$ is admissible in $LTL_{Past,m}$
}.
\end{lemma}

Next necessary for our approach  result  is
the following statement:

\begin{lemma} \label{oox1} {\sl
For any given  rule $\bf r_{nf}$
 in  reduced
normal form, $\bf r_{nf}$  is not admissible in the logic
$LTL_{Past}$  if and only if $\bf r_{nf}$ is refuted in a {\bf special} finite frame
  by a valuation $V$ possessing {\bf some special properties}
(and the size
of the this frame, as earlier, has an effective  bound computable from the
size of $\bf r_{nf}$). }
\end{lemma}


Based at Lemmas \ref{ooxx} and \ref{oox1} we obtain

\begin{theorem} \label{ooxx} {\sl
The logic $LTL_{Past,m}$ is decidable w.r.t. admissible rules.
There is an algorithm verifying for any given inference rule if it is admissible in
$LTL_{Past,m}$}.
\end{theorem}

It is good time to give examples admissible for $LTL_{Past,m}$ but invalid rules.
For example the rules

\[ \nn x / x, \ \ \nn x_1 \ii \nn x_2/ x_1  \ii x_2, \ \
\nn x_1 \ \uu \ \nn x_2 / x_1 \uu x_2   \] are admissible but
invalid in $LTL_{Past,m}$. This is because

\begin{theorem}{\em
If $\varphi(p_1, \dots , p_n, q_1, \dots , q_m)$ is an arbitrary
boolean formula constructed from propositional letters $p_1, \dots ,
 p_n, q_1, \dots , q_m$, then the rule

\[ \frac{ \varphi(\nn x_1, \dots ,  \nn x_n,
 \nn y_1 \uu  \nn z_1 \dots , \nn y_m \uu  \nn z_m )  }
 { \varphi(x_1, \dots ,  x_n,
 y_1 \uu z_1 \dots , y_m \uu z_m ) ) } \]
is admissible in $LTL_{Past,m}$}
\end{theorem}

We may show it exactly the same way as it was proved for the logic $LTL$ itself in \cite{ry08a}.
 This
theorem gives an infinite set of admissible in $LTL_{Past,m}$
rules, were some infinite part of this set consists of
rules invalid in $LTL_{Past,m}$. Actually these rules alow to
withdraw operation $\nn$ from formulas of the premisses.

It is clear that $\Box x \ii \Box\Box x  \ \in \llll$ but $ \Box x \ii \Box\Box x \ \notin LTL_{Past,m}$.
Thus, we immediately see,  there are admissible in $\llll$ rules which are not admissible in $LTL_{Past,m}$.
However even vise versa,

 \begin{theorem} \label{art2} There are not passive inference rules (which means their premisses are unifiable)
 which are admissible
 in  $LTL_{Past,m}$ but not admissible in
 $\llll$.
  \end{theorem}
This statement is already not immediate or trivial.
Thus non-transitive temporal linear logics $\llll$ and $LTL_{Past,m}$
differ w.r.t. admissible inference rules: no one set is enclosed into the another one.

Now we would like to comment the case of temporal linear non-transitive logic
with non-uniform bound of intransitivity. We reported about this logic in \cite{vr14g}.
Just to recall the definition and results,

\begin{definition}
A {\it \bf non-transitive possible-worlds frame}
is 
\[\cm := \langle N, \geq, \mathrm{Next}, \bigcup_{ i \in N} R_i  \rangle,\]
where each $R_i$ is the standard linear order ($\geq$) on the interval $[i,m_i]$, where
$m_i \in N, m_i > i$ and $m_{i+1} > m_i$.
\end{definition}

Again as earlier we may define a model $\cm$ on $\cf$
by introducing a valuation   $V$ on $\cf$
and extend it on all formulas as earlier. In particular,
 for formulas  $ \ppp \suu \psi$:

\begin{definition} Computation rule for {\bf weak, non-unform since}:

\smallskip

$$(\cm,a) \vv_V (\ppp \ \suu \ \psi) \  \ \ \lri  \ \ \  $$

$$ \exists b [
(b R_{a}   a)\wedge ((\cm,b)\vv_V \psi) \wedge
\forall c [(a \leq c < b) \ri
(\cm,c)\vv_V \ppp ]].
 $$
 \smallskip
\end{definition}

\begin{definition}
The logic $LTL_{Past}$ is the set of all formulas which are valid  at any model $\cm$ with any
 valuation.
\end{definition}

The relation
$\bigcup_{ i \in N} R_i $ is again non-transitive and the bounds non-transitivity in frames is arbitrary, - non-uniform.
Using an approach similar to the one used in this paper above we stated the following:

\begin{lemma} \label{oo1} {\sl For any given  rule $\bf r_{nf}$ in  reduced
normal form, $\bf r_{nf}$  is refuted  in a frame of $\cf$ iff $\bf r_{nf}$ can be
refuted in some finite frame $\cf(N^{-})$ by a valuation $V$, where the size
of the frame $\cf(N^{-})$ has size effectively computable from the
size of $\bf r_{nf}$. }
\end{lemma}

And based at this we deducted:

\begin{theorem} \cite{vr14g}\label{bn1} {\it The logic
 $LTL_{Past}$ is
decidable; the satisfiability problem for the non-transitive linear temporal logic
 $LTL_{Past}$ is
decidable: for any formula we can compute if it is satisfiable and
to compute the valuation of the model satisfying this formula.}
\end{theorem}

But at \cite{vr14g} we were not able to solve in $LTL_{Past}$ the problem of admissibility  for inference rules.
Here we solved admissibility problem for $LTL_{Past,m}$, but the presence
of uniform bound $m$ for non-transitivity was impotent and necessary to develop indispensable  technique.

\section{Applications: knowledge of agent's}

In this section we would like to describe applications of our technique and obtained results for formalization the conception of {\em knowledge} in logical terms (we commented at the introduction, that there is a extensive research devoted to study  {\em knowledge} in logical framework, and -- especially epistemic modal logic; here we would suggest somewhat natural and simple, but anyway it seems new).
 We start from a trivial statement
 that
 {\em knowledge}  is not absolute and depends on opinions of individuals (agents) who
accept a statement as safely true or not. Yet, it is not unequally defined what we actually, in fact, consider as knowledge.
First,  would like to look at it via temporal perspective.
Some evident trivial observations are that
\bigskip

{\em
(i) Human beings remember (at least some) past, but

\medskip

(ii) they do not know future at all (rather could surmise what will happen in

immediate proximity time steps);

\medskip

(iii) individual memory tells to us  that the time in past was linear

(though it might be only our perception).

\medskip

}

Therefore it looks meaningful to look for the interpretation of {\em knowledge} in
 past linear temporal logic -  $LTL_{Past}$.
Here below we will use the unary logical operations $K_i$ with meaning - it is a knowledge operation.
So, $K_i \ppp$ says that the statement $\ppp$ is {\em knowledge}.

\medskip

{\bf (i) approach: when knowledge  holds  stable:}

\[ (\nnn^{-},a) \vv_V K_1 \ppp \  \ \ \lri  \ \ \ \exists b [
(b\geq a)\wedge ((\nnn^{-},b)\vv_V \ppp) \wedge  \]

\[ \forall c [(a \leq  c < b) \ri (\nnn^{-},c)\vv_V \ppp ].
\]

 That is

 \[  (\nnn^{-},a) \vv_V K_1 \ppp \ \ \ \lri    \ \ \  (\nnn^{-},a) \vv_V  \ppp S \ppp .\]

 This is an unusual but rather plausible interpretation.
 In time being, we say that $\ppp$ is a knowledge if one day in past it happened
and since then is true until now, today.
The only disturbing point here is that $b$ could be equal $a$ - so, $b$ is today, so then knowledge has very sort time support... Therefore such interpretation yet needs a refining.

\medskip

{\bf (ii) approach: knowledge  if always  was true: }

\[  (\nnn^{-},a) \vv_V K_2 \ppp \ \ \ \lri \ \ \   (\nnn^{-},a) \vv_V \neg ( \top S \neg \ppp) . \]

This is close to standard interpretation in epistemic logic offered quit a while ago:
we consider a fact to be knowledge if it has been true always (in our approach -- in past).

\medskip

{\bf (iii) approach: via parameterized knowledge: }

\[ (\nnn^{-},a) \vv_V K_{\psi} \ppp \ \ \  \lri   \ \ \ (\nnn^{-},a) \vv_V  \ppp S \psi. \]

This means $\ppp$ has a stable value  true, -- since some event happened in past,
which is modeled now by $\psi$ to be true at a state.
Thus, as soon as $\psi$ happened to be true, $\ppp$ always has been true until now.
Clearly this approach generalizes first two suggested above  and yet it is more flexible.
From technical standpoint
we just use standard {\em Since} operation of the linear temporal logic LTL but diverted to past.
 This approach is looking very natural; it gives new view angle on the problem and yet uses some respectful and well established technique.

\medskip

{\bf (iv) approach: via agents knowledge as voted truth for

\ \ \ \ \ the valuation:}

\smallskip

This is very well established area, cf. the book Fagin et al
\cite{fag1}
and more contemporary publications
e.g. -  Rybakov \cite{ry2003,vr179}. Though here we would like to look at it from an another standpoint.
Earlier logical knowledge operations (agents knowledge) were just unary logical operations
$K_i$ interpreted as $S5$-modalities,
 and knowledge operations were introduced via
the vote of agents about truth the statement, etc.
We would like to suggest here somewhat very simple, but it seems natural and new.

Here, in order to implement multi-agent's framework we assume that all agents have theirs own valuations at the frame $N^{-}$.
So, we have $n$ much agents, and $n$-much valuations $V_i$, and as earlier the truth values w.r.t. $V_i$ of any propositional letter $p_j$ at any world $a\in N$.
For applications viewpoint, $V_i$
correspond to agents information about truth of $p_j$ (they may be different).
So, $V_i$ is just individual {\em information} .

How the information can be turned to {\em local} knowledge?
One way is the voted value of truth: we consider a new valuation $V$, w.r.t. which  letters $p_j$ are true at $a$ if majority, biggest part (which one is negotiable) of agents,
believes that $p_j$ is true at $a$. Then we achieve a model with a single
(standard) valuation $V$. Then we can apply {\em any of proposed upper approaches}
to introduce, so to say, logical operations of global knowledge $K$.
If we accept it, our technique works and we may compute true laws of for statements about
knowledge logical operations.

\medskip

{\bf (v) approach: via agents knowledge as conflict resolution:

\ \ \ \ \ at evaluating point}

\smallskip

Here we suggest a way starting similar as in the case (iv) above until introduction of
different valuations $V_i$ of agent's opinion.
But then we suggest

 \[  (\nnn^{-},a) \vv_V K_{\psi} \ppp \ \ \  \lri \ \ \   \forall i [(\nnn^{-}, a)  \vv_{V_i}  K_{\psi} \ppp ]. \]

 In this case, if we will allow nested knowledge operations together with several valuations $V_i$ for agent's information and yet derivative valuation $V$ for all cases when we
 evaluate $K_{\psi}\ppp$ (regardless for which agent (i.e. $V_i$)),
 no decision procedure is known, and our technique does not work.
We think that to solve this open problem it is an interesting open question.


\section{Open problems}

There are many remaining open interesting problems in our suggested framework.
E.g.the problem described at the final part of Section 6 (approach (v)) for multi-agent case.
Yet one more open interesting question is to extend the suggested approach to linear logic based at all integer
numbers $Z$  (which means we have potentially infinite past and infinite future). Then the knowledge will be interpret by a stable truth on reasonably long intervals of time in past and future, so we will need to use both
operations $\suu$ and $\uu$. Next, it would be good to study such approach to the case of continuous time (in past, or both - past and future). Yet the case of admissibility problem for rules in $LTL_{past}$, as well as 
the ones for rules with coefficients
remain open.


\end{document}